\documentclass[manuscript]{aastex}

\slugcomment{Submitted to ApJ Letters}

\shorttitle{Transit and Eclipse of GJ\,436b}
\shortauthors{Deming et al.}

\begin{document}

\title{{\it Spitzer} Transit and Secondary Eclipse Photometry of GJ\,436b}


\author{Drake Deming}
\affil{Planetary Systems Laboratory\\
    Goddard Space Flight Center, Code 693, Greenbelt MD 20771}

\author{Joseph Harrington}
\affil{Department of Physics,\\ Univ. of Central Florida, Orlando, FL 32816}

\author{Gregory Laughlin}
\affil{Department of Astronomy and Astrophysics,\\ 
Univ. of California at Santa Cruz, Santa Cruz CA 95064}

\author{Sara Seager}
\affil{Department of Physics, and Department of Earth, Atmosphere and Planetary Sciences,\\ 
  Massachusetts Institute of Technology, Cambridge MA 02159}

\author{Sarah B. Navarro \& William C. Bowman}
\affil{Department of Physics,\\ Univ. of Central Florida, Orlando, FL 32816}

\and 

\author{Karen Horning}
\affil{Department of Physics \& Space Sciences,\\ 
  Florida Inst. of Technology, Melbourne FL 32901}


\begin{abstract}

We report the results of infrared ($8\,\mu$m) transit and secondary
eclipse photometry of the hot Neptune exoplanet, GJ\,436b using {\it
Spitzer}.  The nearly photon-limited precision of these data allow us
to measure an improved radius for the planet, and to detect the
secondary eclipse.  The transit (centered at $HJD =
2454280.78149\pm0.00016$) shows the flat-bottomed shape typical of
infrared transits, and it precisely defines the planet-to-star radius
ratio ($0.0839 \pm 0.0005$), independent of the stellar properties.
However, we obtain the planetary radius, as well as the stellar mass
and radius, by fitting to the transit curve simultaneously with an
empirical mass-radius relation for M-dwarfs ($M=R$).  We find
$R_{*}=M_{*}=0.47\pm0.02$ in solar units, and $R_{p}=27,600\pm1170$ km
($4.33\pm0.18$ Earth radii).  This radius significantly exceeds the
radius of a naked ocean planet, and requires a gaseous
hydrogen-helium envelope.  The secondary eclipse occurs at phase
$0.587\pm0.005$, proving a significant orbital eccentricity
($e=0.150\pm0.012$).  The amplitude of the eclipse ($5.7\pm0.8 \times
10^{-4}$) indicates a brightness temperature for the planet of
$T=712\pm 36$K.  If this is indicative of the planet's physical
temperature, it suggests the occurrence of tidal heating in the
planet.  An uncharacterized second planet likely provides
ongoing gravitational perturbations that maintain GJ\,436b's orbit
eccentricity over long time scales.

\end{abstract}


\section{Introduction}

The transit and secondary eclipse of an extrasolar planet allow us to
deduce its physical properties to a degree that is not possible in
other observing geometries \citep{charb_ppv}. GJ\,436b \citep{butler}
was recently discovered to be the first transiting Neptune-sized
planet \citep{gillon}, opening new parameter space for exoplanet
studies. It orbits an M-dwarf star lying 10 pc from our solar system
\citep{maness}. In order to constrain the bulk composition and
internal structure of transiting planets \citep{seager, fortney},
precise radii and temperature measurements are needed.  The relatively
small size of GJ\,436A ($\sim 0.4$ solar radii) enhances the
planet-to-star contrast during transit and eclipse. Nevertheless, the
shallow depth of the GJ\,436b transit ($0.006$) is a challenge for
ground-based photometry.  Although ground-based observers are
achieving impressive levels of precision \citep{winn}, photometry from
space-borne platforms remains the preferred observational technique
for the highest-precision transit measurements.  This is especially
true for secondary eclipse, where {\it Spitzer} measurements have been
dominant \citep{charb, deming1, deming2, harrington, knutson}.


In this Letter, we report {\it Spitzer} 8~$\mu$m transit and eclipse
observations of GJ\,436b, and we use these data to refine estimates of
the planet's radius, temperature, and internal structure.

\section{Observations}

The announcement of GJ\,436b transits \citep{gillon} was fortuitously
concurrent with a window of observability using {\it Spitzer}.
Accordingly, we immediately scheduled observations of one transit, and
one secondary eclipse, under our GO-3 Target of Opportunity (ToO)
Program (J.~Harrington, P.I.).  Since the precision required for
measurements of this type is daunting, observations must be carefully
designed to limit instrumental systematics
\citep{harrington}. Moreover, the reported eccentricity of the
GJ\,436b orbit \citep{maness} adds significant uncertainty to the
timing of the secondary eclipse observations. Our community ToO
program thus works with cooperating teams to design observations and
analyze data in line with the best practices gleaned from
experience. We solict collaborations from cooperating teams that
discover suitable targets.

Both observational sequences for GJ\,436 used the IRAC instrument
\citep{irac} in subarray mode, at 8~$\mu$m only.  The transit sequence
consisted of 0.4-second exposures in blocks of 64, obtaining 445
blocks (204 minutes). The secondary eclipse sequence was the same, but
used 780 blocks (356 minutes).  We planned the eclipse observations
based on $10^{4}$ bootstrap trial fits to the Doppler data
\citep{maness}, to define the probability distribution of eclipse
time. Because of the well known ramp-up in the sensitivity of the IRAC
8~$\mu$m detector during long observing sequences \citep{knutson,
harrington}, we offset the transit observations to begin $\sim2$ hours
before transit center.

\section{Data Analysis}

\subsection{Photometry}

Because GJ\,436 is bright at 8~$\mu$m, and the zodiacal background is
weak in comparison, simple aperture photometery attains nearly
photon-limited precision. Our photometry first applies the calibration
information contained in the FITS headers, to convert the signal
levels to electrons.  Within each 64-frame block, we drop the first
frame and the 58th frame, due to known instrumental effects
\citep{harrington, knutson}. We examine the time variation of signal
level for each pixel in the remaining 62 frames, and correct pixels in
frames that are discrepant by $>4\sigma$ to the median value for that
pixel (this removes energetic particle hits). We sum the intensity in
an 8- by 8-pixel square aperture centered on the star in each frame,
including fractional pixels, and sum again over the 62 frames in the
block.  We varied the aperture size to verify that an 8-pixel box
produced the lowest noise, but this dependence is not strong.  We fit
a Gaussian to the peak in a histogram of pixel intensities for each
block to determine, and subtract, the average background level.  We
used the same background value for all 62 frames in a block.

We calculated the expected noise level for the photometry, based on
the Poisson electron counting noise (dominant), and read noise
(small). Comparing the aperture photometry for the 62 frames within
each block, we find that these photometry errors are distributed as
Gaussian noise, with a dispersion merely $3.5\%$ greater than
predicted.  The intensities for the 445 transit blocks are illustrated
in Figure~1, top panel.  We examined the block-to-block variation in
intensity for these points after removing the best transit fit, and we
find a Gaussian distribution, with a standard deviation of $7 \times
10^{-4}$.  Since we detect $\sim 3.1 \times 10^{6}$ electrons per
block (Figure~1), we expect a photon-limited precision of $5.7 \times
10^{-4}$.  We thus attain about $80\%$ of photon-limited S/N,
consistent with previous {\it Spitzer} photometry at this wavelength
\citep{knutson, harrington}.

IRAC photometry at 8~$\mu$m is known to exhibit a gradually increasing
ramp-up in sensitivity, due to filling of charge traps in the
detectors \citep{knutson, harrington}.  This ramp is visible in the
top panel of Figure~1, but is weaker than usual for the transit data
(the ramp varies due to prior usage of the detector). We removed it by
masking out the data near transit, and fitting a parabola to the
out-of-transit points.  We have considerable experience in fitting to
this ramp, via our monitoring of GJ\,876 (program 30498).  Even strong
ramps can be fit by the sum of a linear plus logarithmic function,
using linear regression.  We applied this more elaborate procedure to
the GJ\,436 transit ramp, but found no significant difference with the
simple parabola fit. In the case of the secondary eclipse (Figure~2),
the ramp is stronger, but is still well removed by our full linear +
logarithmic fit. We conclude that this ramp is properly reproduced in
both cases, and does not contribute significantly to our errors.

\subsection{Transit Parameters}

A feature of IR transit measurements is the virtual lack of stellar limb
darkening.  Not only does this produce a simple box-like shape for the
transit, but \citet{richardson} suggest that it can increase the radius
precision for a given level of photometric precision. Our analysis
adopts the (small) limb darkening for GJ\,436A based on a Kurucz model
atmosphere for 3500/5.0/0.0 in Teff/log(g)/[M/H]. We verified that
changing the stellar temperature, gravity, or metallicity within the
errors \citep{maness, bean} has negligible effect, because the limb
darkening remains small over the plausible range.  We integrated the stellar
center-to-limb intensities in the Kurucz model over the bandpass of
the IRAC 8~$\mu$m filter to obtain the limb darkening appropriate to
this IR transit.  Since this small IR limb darkening is not included
in the \citet{claret} prescriptions, we generate theoretical transit
curves numerically.

We compute theoretical transit curves by tiling the star in a
latitude-longitude grid with zone spacing of $0.18$ degrees, and
applying the IRAC 8~$\mu$m limb darkening.  We pass the planet across
the numerical star in steps of 0.01 stellar radii, with the planet
radius and impact parameter specified in units of the stellar
radius. To increase precision, stellar zones at the edge of the planet
are adaptively sub-sampled in a $10 \times 10$ finer grid.  We
verified the code's precision (better than $10^{-6}$) by comparing to
the \citet{mandel} analytic non-linear limb darkening cases, and by
comparing the depth of synthetic transits to the planet-to-star area
ratio (${R_p}^2/{R_*}^2$), for the case when limb darkening is
identically zero.

Fitting to high precision transit photometry requires a determination,
or assumption, of the stellar mass \citep{brown}.  \citet{gillon}
adopted $0.44$ solar masses for GJ\,436A, based on the observed
luminosity, and they cited the empirical M-dwarf mass-radius relation
from \citet{ribas} ($R=M$ in solar units), to justify $0.44$ solar
radii for the stellar size. Our fit procedure is somewhat different.
Given the lack of limb darkening, we can immediately determine the
ratio of planet to stellar radius as $0.0839 \pm 0.0005$ from the
depth of the transit (Figure~1).  With this value fixed, we generate a
grid of transit curves for a range of impact parameters.  At each
impact parameter, we vary the adopted stellar mass, and compute the
transverse velocity of the planet across the star. This computation
uses the orbital elements from a fit to the Doppler data
\citep{maness}, constrained by the secondary eclipse time (see
below). We vary the stellar radius to convert the radius increments on
the abscissa of the synthetic transit curve to orbit phase, using the
calculated transverse velocity. We include a shift in phase for the
synthetic transit curve, to allow for imprecision in the
\citet{gillon} ephemeris. In this manner, we find the best fit stellar
radius versus stellar mass, and a revised transit time.  We estimate
the stellar radius precision from the variation in ${\chi}^2$ at a
given mass.

We intersect the radius versus mass relation ($R \sim M^{0.33}$) from
the fitting procedure with the empirical mass-radius relation $R=M$
\citep{ribas} to find the best stellar mass and radius, and planet
radius, at each impact parameter.  Repeating this over a grid of
impact parameters, we adopt the best fit from the global minimum
${\chi}^2$.  We determine the error range from $\delta{\chi}^2$, and
from visually inspecting the quality of the fits, paying particular
attention to ingress/egress.  Our fitting always uses the unbinned
data (Figure~1, top), but we bin the data for the lower panel of
Figure~1, to better illustrate the quality of the fit.  The
derived time of transit center is $HJD = 2454280.78149\pm0.00016$.

\subsection{Secondary Eclipse}

The secondary eclipse is shown in Figure~2.  The top panel plots the
bulk of the data (omitting some points at the outset); the eclipse
occurs near the end of the observational sequence, at phase
$0.587\pm0.005$, with amplitude $5.7\pm0.8 \times 10^{-4}$ in units of
the stellar intensity.  Like the transit, all of our fits to this
event were made on the original, unbinned data.  However, for clarity,
the lower panel of Figure~2 shows binned data, expands the phase
scale, and overplots the best fit eclipse curve. Our fit constrains
the duration of eclipse to equal the duration of transit, finding only
the amplitude and central phase.

Table~1 summarizes our results for transit and secondary eclipse.

\section{Results and Discussion}

Our result for the stellar mass and radius is $M=R=0.47\pm0.02$ in
solar units. We are encouraged that these are close to values ($0.44$)
constrained by independent data \citep{maness}.  Our derived planet
radius is $R_{p}=27,600\pm1170$ km ($4.33\pm0.18$ Earth radii). We
conclude that this planet is larger than originally indicated by
ground-based photometry \citep{gillon}.  Since this radius is
significantly larger than all planets of exclusively solid composition
\citep{seager, fortney}, GJ\,436b must have a significant, gaseous, 
hydrogen-helium envelope.  After our transit analysis was complete, we
became aware of an independent analysis of the transit data (but not
the eclipse data) by \citet{gillon2}.  These authors do not vary the
stellar mass in their fit, but they obtain a very similar radius for this
planet, and arrive at essentially the same conclusion.

The observed phase of the secondary eclipse, $\phi=0.587\pm 0.005$
indicates that the orbit of GJ\,436b is significantly
eccentric. Assuming a longitude of pericenter $\varpi=0$, the
magnitude of the observed timing offset indicates a minimum orbital
eccentricity, $e_{\rm min}=0.137 \pm 0.007$.

Using the constraints provided by the observed times of central
transit, $T_{c}=2454222.616$ HJD \citep{gillon}, our
$T_{c}=2454280.78149$ HJD, and the observed secondary eclipse at
$T_{s}=2454282.33 \pm 0.01$ HJD, we obtained a set of single-planet
Keplerian fits to the radial velocity data published by
\citet{maness}. A straightforward bootstrap resampling procedure
\citep{press} yields $e=0.150\pm 0.012$, $\varpi=343\pm14^{\circ}$,
and $M=0.070 \pm 0.003\, M_{\rm Jup}$.

As a consequence of its non-zero orbital eccentricity, GJ\,436b is
likely experiencing asynchronous rotation. \citet{hut} gives an
expression for the spin period of a zero-obliquity spin pseudo-synchronized planet:
\begin{equation}
P_{\rm spin}={ (1+3e^2 + {3\over{8}}e^4)(1-e^2)^{3/2} \over
{1 + {15\over{2}} e^2+ {45\over{8}}e^4+ {5\over{16}}e^6}} P_{\rm orbit}
\end{equation}
For GJ\,436b, we find $P_{\rm spin}=2.32 \, {\rm d}$, which yields a
19-day synodic period for the star as viewed from a fixed longitude on
the planet. The large orbital eccentricity also indicates that a
significant amount of tidal heating must be occurring. To second order
in eccentricity, the tidal luminosity of a spin-synchronous planet
\citep{peale, mardling} is given by:
\begin{equation}
{dE\over{dt}}={21\over{2}}{k_{2}\over{Q}}{GM_{\star}^{2} n {R_{\rm
p}^{5}}e^{2} \over{a^6}}
\end{equation}
where $k_2$ is the planetary potential Love number of degree 2, $n$ is
the orbital mean motion, $a$ is the orbital semimajor axis, and $Q$ is
the planet's effective tidal dissipation parameter. The analysis of
\citet{levrard} indicates that the tidal luminosity of an
asynchronously rotating planet with $e\sim0.15$ will exceed the value
implied by the above expression by a small amount.

If we adopt $T_{\rm eff}=3350 K$ for GJ\,436A, take a zero albedo for
the planet, and assume a uniform re-radiation of heat from the entire
planetary surface, we obtain a planetary $T_{eq}=642\,{\rm K}$. The
somewhat higher temperature ($T=712\pm 36$K) implied by the secondary
eclipse depth could arise from inefficient transport of heat to the
night side of the planet, from a non-blackbody planetary emission
spectrum, from tidal luminosity, or from a combination of the three.

We can estimate $Q$ by assuming $A=0$, uniform re-radiation, and a
blackbody planetary SED to find a fiducial tidal luminosity of
$4.7\times10^{26} \, {\rm ergs sec^{-1}}$, and
${Q\over{k_2}}=2.1\times10^{4}$. Assuming $k_2=0.34$ (the Jovian
value), gives $Q=7140$. This value is in rough accord with the
$Q$-values measured for Uranus and Neptune.  Banfield and Murray
(1992) derive $1.2\times10^{4}<Q_{N}<3.3\times10^{5}$ for Neptune,
whereas Tittemore and Wisdom (1989) employ the Uranian satellites to
derive $Q_{U}<3.9\times10^{4}$.  If the planet has maintained
$e\sim0.15$ for billions of years, $Q\sim7000$ indicates that the
planet has radiated tidal energy comparable to the orbital energy and
in excess of 100 times its own gravitational binding
energy. 

Furthermore, $Q\sim7000$ implies a circularization timescale
${e\over{{de/dt}}}\sim3.0\times10^{7}$\,yr. Indeed for any range of
$Q$ - which is uncertain even for solar system bodies - the
circularization timescale is $<10^{8}$ years. It is thus
highly likely that an as-yet uncharacterized second planet is
providing ongoing gravitational perturbations that allow GJ\,436b's
eccentricity to be maintained over long timescales.


\acknowledgments

We are grateful to the staff at the Spitzer Science Center for their
prompt and efficient scheduling of our observations.



{\it Facilities:} \facility{Spitzer}.




\begin{figure}
\epsscale{0.6}
\plotone{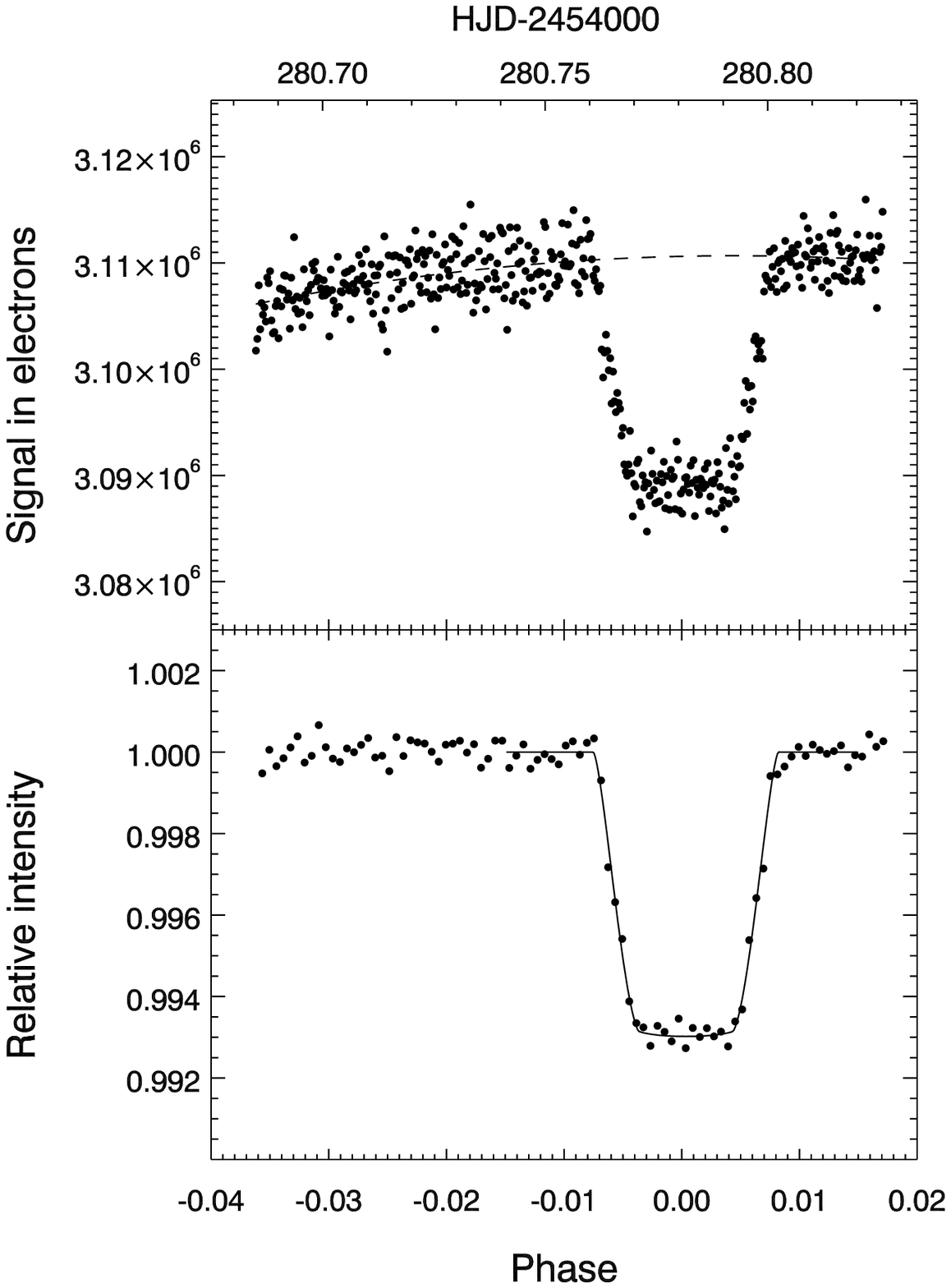}
\caption{Spitzer photometry of the GJ\,436b transit with fitted
transit curves.  Top: photometry before baseline correction. The
dashed line is the adopted baseline. Bottom: Baseline-corrected data,
binned to approximately 2-minute time resolution ($137$ sec), with the
best fit transit curve. Note the flat bottom that proves a non-grazing transit.
 \label{fig1}}
\end{figure}

\clearpage

\begin{figure}
\epsscale{0.6}
\plotone{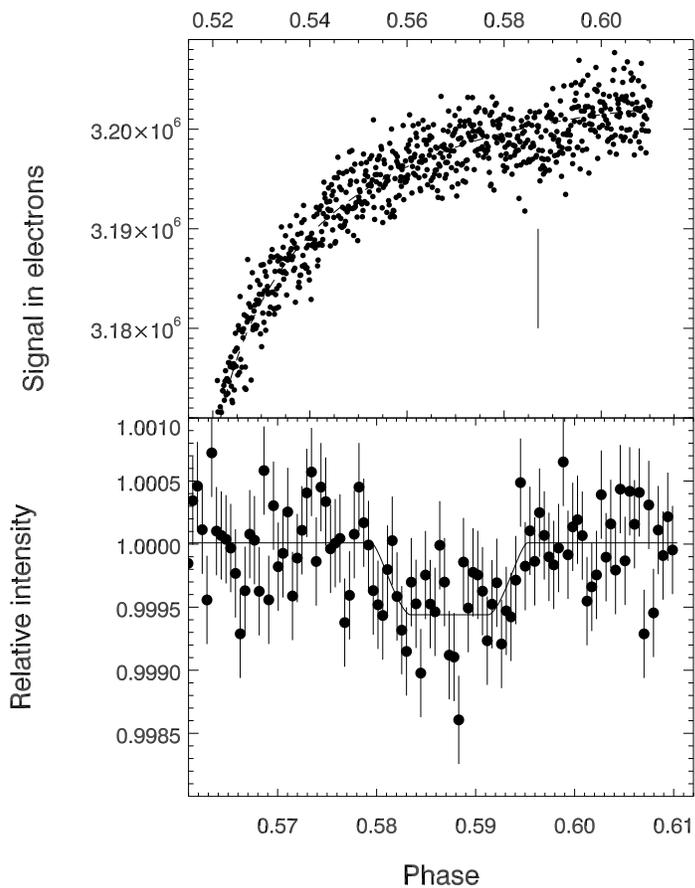}
\caption{Secondary eclipse photometry of GJ\,436b.  Top: photometry
showing the ramp in intensity, with the eclipse marked at phase
$0.587$. The dashed line is the adopted baseline for points with phase
$>0.52$. Bottom: Binned data ($102$-sec time resolution) shown in
comparison to the best fit secondary-eclipse curve, whose amplitude is
$5.7 \pm 0.8 \times 10^{-4}$.}
\end{figure}

\clearpage

\begin{deluxetable}{ll}
\tabletypesize{\scriptsize}
\tablecaption{Derived parameters for GJ\,436.}
\tablewidth{0pt}
\tablehead{
\colhead{Parameter} & \colhead{Value}
}
\startdata
Stellar radius\tablenotemark{a} & $0.47\pm 0.02$  \\
Stellar mass & $0.47\pm 0.02$ \\
Planet radius & $27,600\pm1170$~km \\
Impact parameter & $0.85^{+0.03}_{-0.02}$ \\
Transit time & $HJD=2454280.78149\pm0.00016$  \\
Orbit semi-major axis & $0.0291\pm 0.0004$ AU \\
Orbit eccentricity & $0.150\pm 0.012$ \\
$a/R_{*}$ & $13.2\pm 0.6$ \\
Secondary eclipse phase & $0.587\pm0.005$ \\
Secondary eclipse amplitude & $5.7\pm0.8 \times 10^{-4}$ \\
Planet brightness temperature & $712\pm36$\,K \\
Planet mass &  $0.070 \pm 0.003\, M_{\rm Jup}$ \\
\enddata
\tablenotetext{a}{Radius constrained to equal mass, in solar units.}
\end{deluxetable}

\end{document}